\documentstyle[aps,twocolumn]{revtex}
\newcommand{\be}{\begin{equation}}
\newcommand{\ee}{\end{equation}}
\newcommand{\bea}{\begin{eqnarray}}
\newcommand{\eea}{\end{eqnarray}}
\newcommand{\dd}{\dagger \hspace{-0.7mm} \dagger}
\newcommand{\venus}{ {\scriptscriptstyle +} \hspace{-1.5mm}^{\circ}}
\newcommand{\aeq}{&=&}

\newcommand{\itPi}{{\it \Pi}}
\newcommand{\itSigma}{{\it \Sigma}}

\newcommand{\bra}{\langle}
\newcommand{\ket}{\rangle}
\newcommand{\dbra}{\bra \! \bra}
\newcommand{\dket}{\ket \! \ket}

\newcommand{\me}{\mbox{e}}

\newcommand{\strat}{\stackrel{\circ}{,}}

\begin{document}

%%%%%%%%%%%%%%%%%%%%%%%%%%%%%%%%%%%%%%%%%%%%%%%%%%%%%%%
% Title etc.
%%%%%%%%%%%%%%%%%%%%%%%%%%%%%%%%%%%%%%%%%%%%%%%%%%%%%%%
\title{QUANTUM STOCHASTIC DIFFERENTIAL EQUATIONS IN VIEW OF
NON-EQULIBRIUM THERMO FIELD DYNAMICS
%\thanks{A talk provided 
%for the Conference on {\it Gaussian Process and Quantum Analysis}
%held at Nagoya University in Japan during the period of 
%October 29--31, 1998.}
}

\author{Toshihico Arimitsu\thanks{arimitsu@cm.ph.tsukuba.ac.jp}}

\address{Institute of Physics, University of Tsukuba,
Ibaraki 305-8571, Japan}

\date{\today}

\maketitle

\begin{abstract}

Most of the mathematical approaches for quantum Langevin equation
are based on the non-commutativity of the random force operators.
Non-commutative random force operators are introduced 
in order to guarantee that the equal-time commutation relation for 
the stochastic annihilation and creation operators preserves in time.
If it is true, it means that the origin of dissipation is of
quantum mechanical. However,  physically, it is hard to believe it.
By making use of the unified canonical operator formalism
for the system of the quantum stochastic differential equations 
within Non-Equilibrium Thermo Field Dynamics, it is shown that
it is not true in general.

\end{abstract}

\pacs{}

\narrowtext

%%%%%%%%%%%%%%%%%%%%%%%%%%%%%%%%%%%%%%%%%%%%%%%%%%%%%%%%%%%%%%%%%%%
% Text
%%%%%%%%%%%%%%%%%%%%%%%%%%%%%%%%%%%%%%%%%%%%%%%%%%%%%%%%%%%%%%%%%%%

\section{Introduction}

The studies of the Langevin equation for quantum systems
were started in connection with the development of laser
\cite{Senitzky,Lax,Haken}, and are still continuing in order to
develop a satisfactory formulation 
\cite{Streater,Hasegawa,Accardi,Hudson,Hudson85,Parth}
(see comments in \cite{Kubo1969}).
Most of the mathematical approaches for quantum Langevin equation
are based on the non-commutativity of the random force operators.
For dissipative systems, for example, 
we have equations for the operators
$\bra a(t) \ket$ and $\bra a^\dagger(t) \ket$ 
averaged with respect to random force operators of the forms
\bea
\frac{d}{dt} \bra a(t) \ket \aeq -i\omega \bra a(t) \ket
-\kappa \bra a(t) \ket,
\\
\frac{d}{dt} \bra a^\dagger(t) \ket \aeq i\omega 
\bra a^\dagger(t) \ket
-\kappa \bra a^\dagger(t) \ket,
\eea
with the initial condition
\be
\bra a(0) \ket = a, \qquad \bra a^\dagger(0) \ket = a^\dagger,
\ee
where $a$ and $a^\dagger$ satisfy the canonical commutation relation
\be
[a,\ a^\dagger] =1.
\label{comm rel}
\ee
The equal-time commutation relation for these operators 
decays in time:
\be
[ \bra a(t) \ket,\ \bra a^\dagger(t) \ket] = \me^{-2\kappa t}.
\ee
Random force operators $df(t)$ and $df^\dagger(t)$ are introduced 
in order to rescue this situation.
If the random force operators in the Langevin equations
\bea
d a(t) \aeq -i\omega a(t) dt -\kappa a(t) dt + \sqrt{2\kappa} \ df(t),
\label{Lan a}
\\
d a^\dagger(t) \aeq i\omega a^\dagger(t) dt 
-\kappa a^\dagger(t) dt + \sqrt{2\kappa} \ df^\dagger(t),
\label{Lan a dag}
\eea
satisfy
\be
[df(t),\ df^\dagger(t) ] = dt,
\ee
the equal-time commutation relation for 
the stochastic operators $a(t)$ and $a^\dagger(t)$ preserves in time:
\be
d \left( [a(t),\ a^\dagger(t)] \right) =0,
\ee
meaning that
\be
[a(t),\ a^\dagger(t)] = 1,
\ee
with (\ref{comm rel}).

The above argument is of zero temperature related only to 
the zero-point fluctuation. However, it 
has been extended to include the situations for finite temperature.
Then, we have a crucial question. Should we interpret that 
the origin of thermal dissipation is quantum mechanical?
In this paper, we will investigate this question with the help of
the system of the stochastic differential equations 
within Non-Equilibrium Thermo Field Dynamics (NETFD) 
\cite{netfd1,netfd2,netfd3,guida1,guida2,can1,can2,jim,kinetic,%
tft1,hydro,essay,proceedings,stoch,Saito,Zubarev memorial,Saito Thesis,%
Imagire,Sudo,Tominaga-Ban,Ban,Tominaga,Iwasaki,Willeboordse,Naoko,%
non-linear,cloud chamber,Kramers eq}.

NETFD is a {\it canonical operator formalism} of 
quantum systems in far-from-equili\-brium state which provides us 
with a unified formulation for dissipative systems 
by the method 
similar to the usual quantum field theory that
accommodates the concept of the dual structure in the
interpretation of nature, i.e.\ in terms of 
the {\it operator algebra} and the {\it representation space}.
The representation space of NETFD (named {\it thermal space}) 
is composed of the direct product of two
Hilbert spaces, the one for {\it non-tilde} fields and the other
for {\it tilde} fields.
It can be said that NETFD is a framework which gives a foundation of 
Green's function formalisms, such as Schwinger's closed-time path
method, Keldysh-method, and so on \cite{schwinger,keldysh,su}, in terms of 
dissipative quantum field operators within the representation space
constructed on an unstable vacuum.

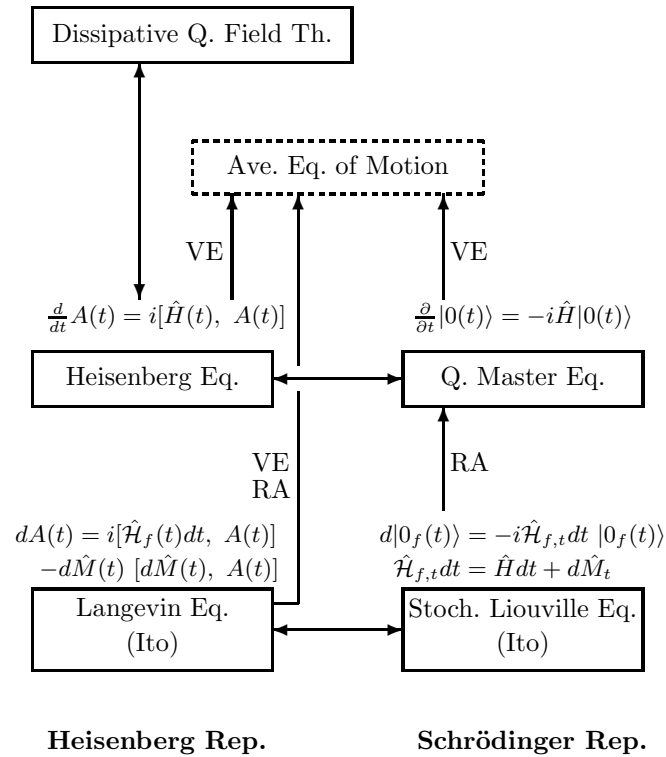
\begin{figure}[h,t,b,p]

\begin{picture}(50,320)(-50,-20)

%\newlength{\gnat}
%\setlength{\gnat}{1cm}

\thicklines

\put(-45,-10){{\bf Heisenberg Rep.}}
\put(95,-10){{\bf Schr\"odinger Rep.}}

\put(-50,20){\framebox(90,30){}}
\put(-50,22){\makebox(90,15){(Ito)}}
\put(-50,35){\makebox(90,15){Langevin Eq.}}
\put(-40,65){\makebox(65,12.5){\small
$
dA(t) = i [ \hat{{\cal H}}_f(t) dt,\ A(t) ] 
$}}
\put(-30,52){\makebox(55,12.5){\small
$
 - d\hat{M}(t)\  [ d\hat{M}(t),\ A(t) ] 
$}}

\put(90,20){\framebox(90,30){}}
\put(90,22){\makebox(90,15){(Ito)}}
\put(90,35){\makebox(90,15){Stoch.\ Liouville Eq.}}
\put(80,65){\makebox(110,12.5){\small
$
d \vert 0_f(t) \ket = -i \hat{{\cal H}}_{f,t} dt \ \vert 0_f(t) \ket
$}}
\put(80,52){\makebox(95,12.5){\small
$
\hat{{\cal H}}_{f,t} dt = \hat{H} dt + d \hat{M}_t
$}}

\put(90,120){\framebox(90,20){Q.\ Master Eq.}}
\put(80,145){\makebox(110,17.5){\small
$
\frac{\partial}{\partial t} \vert 0(t) \ket = -i \hat{H} \vert 0(t) \ket
$}}

\put(-50,120){\framebox(90,20){Heisenberg Eq.}}
\put(-20,145){\makebox(40,17.5){\small
$
\frac{d}{dt} A(t) = i [\hat{H}(t),\ A(t)]
$}}

\put(10,200){\dashbox{2.0}(110,20){Ave.\ Eq.\ of Motion}}

\put(-50,250){\framebox(120,20){Dissipative Q.\ Field Th.}}

\put(105,80){\vector(0,1){40}}
\put(107.5,95){RA}

\put(50,45){\line(0,1){80}}
\put(50,135){\vector(0,1){65}}
\put(50,45){\line(-1,0){10}}
\put(32.5,95){VE}
\put(32.5,85){RA}

\put(40,35){\vector(1,0){50}}
\put(90,35){\vector(-1,0){50}}

\put(105,160){\vector(0,1){40}}
\put(107.5,175){VE}

\put(25,160){\vector(0,1){40}}
\put(7.5,175){VE}

\put(90,130){\vector(-1,0){50}}
\put(40,130){\vector(1,0){50}}

\put(-10,160){\vector(0,1){90}}
\put(-10,250){\vector(0,-1){90}}

\end{picture}

\caption{System of the Stochastic Differential 
Equations within Non-Equilibrium Thermo Field Dynamics. 
RA stands for the random average. 
VE stands for the vacuum expectation.
}
\label{structure}
\end{figure}

In the extension to take 
account of the quantum stochastic processes 
\cite{proceedings,stoch,Saito,Zubarev memorial,Saito Thesis,Imagire},
NETFD again allowed us to construct a unified canonical theory 
of quantum stochastic operators.  The stochastic Liouville
equations both of the Ito and of the Stratonovich types
were introduced in the Schr\"odinger representation. 
Whereas, the Langevin equations both of the Ito and of the
Stratonovich types were constructed as the Heisenberg
equation of motion with the help of the time-evolution generator 
of corresponding stochastic Liouville equations (Fig.\ \ref{structure}).  
The Ito formula was generalized for quantum systems.

NETFD has been applied to various systems, e.g.\
the dynamical rearrangement of thermal vacuum in 
superconductor \cite{Sudo}, spin relaxation \cite{Tominaga-Ban},
various transient phenomena in quantum optics 
\cite{Ban,Tominaga,Iwasaki,Willeboordse,Naoko}, 
non-linear damped harmonic oscillator \cite{non-linear},
the tracks in the cloud chamber 
(a non-demolition continuous measurement) \cite{cloud chamber},
microscopic derivation of the quantum Kramers equation 
\cite{Kramers eq}.

In the next section, the framework of NETFD will be briefly explained. 
In section \ref{two systems}, two systems of the stochastic
differential equations will be introduced, one with {\it non-unitary} 
time-evolution generator and the other with {\it unitary} 
time-evolution generator. Both systems are constructed to be 
consistent with the same quantum master equation.
The key is the existence of the fluctuation-dissipation theorem
between the multiple of martingale operators and the {\it imaginary}
part of hat-Hamiltonian. 
In section \ref{damped oscillator system}, two systems will be applied 
to the model of damped harmonic oscillator interacting with 
irrelevant random force system by the linear dissipative coupling. 
It will be shown that both systems are consistently applicable.
The existence of the non-commutative random force operators is 
essential for the system with unitary time-evolution generator.
In section \ref{Kramers system}, two systems will be applied 
to the model of damped harmonic oscillator interacting with 
irrelevant system by the position-position interaction.
It will be shown that the system with unitary time-evolution 
generator cannot produce the framework which is consistent with
the master equation, since there appear only commutative random force
operators in martingale.
Section \ref{summary} will be devoted to summary and discussion.

\section{Framework of NETFD}

The dynamics of physical systems is described, within
NETFD, by the Schr\"odinger equation for 
the thermal ket-vacuum $\vert 0(t) \ket$:
\be
\frac{\partial}{\partial t} \vert 0(t) \ket 
= -i \hat{H} \vert 0(t) \ket.
\label{Schr eq}
\ee
The time-evolution generator $\hat{H}$ is an tildian operator
satisfying
\be
( i \hat{H} )^\sim = i \hat{H}.
\label{H hat tildian}
\ee
The {\it tilde conjugation} $\sim$ is defined by 
\bea
(A_1A_2)^{\sim} \aeq \tilde{A}_1\tilde{A}_2,\\ 
(c_1A_1+c_2A_2)^{\sim} \aeq c^{*}_1\tilde{A}_1+c^{*}_2\tilde{A}_2,
\\
(\tilde{A})^{\sim} \aeq A,\\ 
(A^{\dagger})^{\sim} \aeq \tilde{A}^{\dagger},
\eea
where $c_1$ and $c_2$ are $c$-numbers.
The tilde and non-tilde operators at an equal time are
mutually commutative:
\be
[A,\ \tilde{B}]=0.
\label{commutable}
\ee

The thermal bra-vacuum $\bra 1 \vert $ is the eigen-vector
of the hat-Hamiltonian $\hat{H}$ with zero eigen-value:
\be
\bra 1 \vert \hat{H} = 0.
\label{left zero}
\ee
This guarantees the conservation of the inner product 
between the bra and ket vacuums in time:
\be
\bra 1 \vert 0(t) \ket = 1.
\label{normalization}
\ee

Let us assume that the thermal vacuums satisfy
\be
\bra 1 \vert^\sim = \bra 1 \vert, \quad 
\vert 0(t_0) \ket^\sim = \vert 0(t_0) \ket,
\ee
at a certain time $t = t_0$. 
Then, (\ref{H hat tildian}) guarantees that 
they are satisfied for all the time:
\be
\bra 1 \vert^\sim = \bra 1 \vert, \quad 
\vert 0(t) \ket^\sim = \vert 0(t) \ket.
\ee

The tilde operator and the non-tilde operator are related by 
the thermal state condition for the bra vacuum:
\be
\bra 1 \vert \tilde{A} = \bra 1 \vert A^\dagger,
\label{thermal state cond}
\ee
which reduces the numbers of the degrees of freedom to 
the original ones. The numbers of the degrees of 
freedom were doubled by the introduction of tilde operators.

The observable operator $A$ should be an Hermitian operator
consisting only of non-tilde operators.

\section{Two Systems of Stochastic Differential Equations}
\label{two systems}

\subsection{Quantum Master Equation}

Let us consider the system of quantum stochastic differential 
equations which is constructed to be consistent with 
the quantum master equation (the quantum Fokker-Planck equation)
\be
\frac{\partial}{\partial t} \vert 0(t) \ket 
= - i \hat{H} \vert 0(t) \ket,
\label{master eq}
\ee
with the hat-Hamiltonian 
\be
\hat{H} = \hat{H}_S + i \hat{\itPi},
\ee
where 
\be
\hat{H}_S = H_S - \tilde{H}_S,
\ee
with $H_S$ being the Hamiltonian of a relevant system.
It is easily seen that $\hat{H}_S$ satisfies
\be
\bra 1 \vert \hat{H}_S = 0.
\ee
It is assumed that the {\it imaginary} part $\hat{\itPi}$ 
of the hat-Hamiltonian can be
divided into two parts, i.e., the relaxational part $\hat{\itPi}_R$
and the diffusive part $\hat{\itPi}_D$:
\be
\hat{\itPi} = \hat{\itPi}_R + \hat{\itPi}_D,
\ee
and each of them satisfies
\be
\bra 1 \vert \hat{\itPi}_R = 0, \quad
\bra 1 \vert \hat{\itPi}_D = 0.
\ee

Introducing the time-evolution operator $\hat{V}(t)$ by
\be
\frac{d}{dt} \hat{V}(t) = - i \hat{H} \hat{V}(t),
\label{V}
\ee
with the initial condition $\hat{V}(0) =1$, we can
define the Heisenberg operator
\be
A(t) = \hat{V}^{-1}(t) A \hat{V}(t),
\ee
which satisfies the Heisenberg equation 
\be
\frac{d}{dt} A(t) = i [\hat{H}(t),\ A(t)],
\label{Heisenberg eq for semi-free}
\ee
for dissipative systems.

The equation of motion for the averaged quantity 
$\bra 1\vert A(t) \vert 0 \ket$ is 
derived by means of the Heisenberg
equation (\ref{Heisenberg eq for semi-free})
by taking its vacuum expectation:
\be
\frac{d}{dt} \bra 1 \vert A(t) \vert 0 \ket = i 
\bra 1 \vert [ \hat{H}(t),\ A(t) ] \vert 0 \ket.
\label{ave eq for semi-free}
\ee
The same equation can be also derived with the help of 
the master equation (\ref{master eq}) as
\bea
\frac{d}{dt} \bra 1\vert A \vert 0(t) \ket \aeq -i 
\bra 1\vert A \hat{H} \vert 0(t) \ket.
\eea

We would like to emphasize here that the existence of the
Heisenberg equation of motion (\ref{Heisenberg eq for semi-free})
for coarse grained operators is one of the notable features
of NETFD. This enabled us to construct a {\it canonical formalism of 
the dissipative quantum field theory}, where the coarse grained 
operators $a(t)$ etc.\ in the Heisenberg representation preserve
the equal-time canonical commutation relation 
\be
[ a(t),\ a^\dagger(t)] = 1, \qquad [ \tilde{a}(t),\ 
\tilde{a}^\dagger(t)] = 1.
\label{commu-c}
\ee

Note that we have an equation of motion for a vector
$\bra 1 \vert A(t)$:
\bea
\frac{d}{dt} \bra 1 \vert A(t)  \aeq i 
\bra 1 \vert [ \hat{H}(t),\ A(t) ]
\nonumber\\
\aeq i \bra 1\vert [H_S(t),\ A(t)]
\nonumber\\
&& - \kappa \left\{ \bra 1\vert [A(t),\ a^\dagger(t)] a(t) 
\right. \nonumber\\
&& \left. 
+ \bra 1\vert a^\dagger(t) [a(t),\ A(t)] \right\}
\nonumber\\
&& + 2\kappa \bar{n} \bra 1\vert [a(t),\ [A(t),\ a^\dagger(t)]]
\label{eq for vector state}
\eea
in terms of only non-tilde operators with the help of the 
condition (\ref{initial bra}).
Applying the ket-vacuum $\vert 0 \ket$ to (\ref{eq for vector state}),
we obtain the equation of motion for the averaged quantity
(\ref{ave eq for semi-free}).

\subsection{Non-Unitary Time-Evolution}

The system of stochastic differential equations with 
{\it non-unitary}
time-evolution is constructed by the following general 
procedures.

The stochastic Liouville equation 
\be
d \vert 0_f(t) \dket = - i \hat{{\cal H}}_{f,t} dt
\vert 0_f(t) \dket,
\label{stoch Liouville Ito}
\ee
of the Ito type is specified with the stochastic hat-Hamiltonian
\bea
\hat{{\cal H}}_{f,t} dt \aeq \hat{H} dt + d \hat{M}_t
\nonumber\\
\aeq \hat{H}_S dt + i \hat{\itPi} dt + d \hat{M}_t,
\eea
where $\hat{\itPi}=\hat{\itPi}_R + \hat{\itPi}_D$ is the same that
appeared in the master equation (\ref{master eq}).
The martingale operator $d \hat{M}_t$ annihilates 
the bra-vacuum $\bra 1 \vert$ of the relevant system:
\be
\bra 1 \vert d \hat{M}_t = 0,
\label{condition for non-unitary martingale}
\ee
which means that the stochastic Liouville equation 
(\ref{stoch Liouville Ito}) preserves its probability 
just within the relevant system. This feature is the same 
as the one within the system of stochastic differential equations 
for classical systems.
The martingale operator satisfies the fluctuation-dissipation 
theorem of the second kind:
\be
d \hat{M}_t d \hat{M}_t = -2 \hat{\itPi}_D dt,
\ee
which should be interpreted as a weak relation.\footnote{
It is similar to the classical cases where 
the fluctuation-dissipation theorem of 
the second kind is specified within the stochastic limit.
}

Applying $\bra \vert$ to (\ref{stoch Liouville Ito}), 
we have an equation for 
\be
\vert 0(t) \ket = \bra \vert 0_f(t) \dket,
\ee
which is nothing but the quantum master equation 
(\ref{master eq}).

Introducing the stochastic time-evolution operator
\be
d \hat{V}_f(t) = - i \hat{{\cal H}}_{f,t} dt
\hat{V}_f(t),
\ee
we can define the stochastic Heisenberg operator
\be
A(t) = \hat{V}_f^{-1}(t) A \hat{V}_f(t),
\ee
which satisfies the stochastic Heisenberg equation
(the Langevin equation)
\bea
dA(t) \aeq i [ \hat{{\cal H}}_f(t) dt,\ A(t) ] 
\nonumber\\
&&- d\hat{M}(t)\  [ d\hat{M}(t),\ A(t) ],
\eea
of the Ito type.
Here, we introduced the martingale operator in the 
Heisenberg representation by
\be
d \hat{M}(t) = d \left( \hat{V}_f^{-1}(t) M_t \hat{V}_f(t) \right).
\ee
Note that
\be
d \hat{M}(t) = d' \hat{M}(t),
\ee
with
\be
d' \hat{M}(t) =  \hat{V}_f^{-1}(t) d M_t \hat{V}_f(t).
\ee

Making use of the relation of the Ito-Stratonovich stochastic
calculus (see Appendix~\ref{Multiplication}), we can derive 
from (\ref{stoch Liouville Ito}) the stochastic Liouville equation 
\be
d \vert 0_f(t) \dket = - i \hat{H}_{f,t} dt \circ
\vert 0_f(t) \dket,
\ee
of the Stratonovich type, where the symbol $\circ$ indicates
the Stratonovich stochastic multiplication.
The stochastic hat-Hamiltonian
\be
\hat{H}_{f,t} dt = \hat{H}_S dt + i \hat{\itPi}_R dt
+ d \hat{M}_t,
\ee
contains only the relaxational part $\hat{\itPi}_R$.

With this hat-Hamiltonian, we can write down the stochastic
Heisenberg equation 
\be
dA(t) = i [ \hat{H}_f(t) dt \strat \ A(t) ],
\ee
of the Stratonovich type.
Note that it does not have the term producing diffusive 
time-evolution, which is the same characteristics that 
appeared in the system of classical stochastic differential
equations.

\subsection{Unitary Time-Evolution}

The system of stochastic differential equations with 
{\it unitary}
time-evolution is constructed by the following general 
procedures.

The stochastic Liouville equation
\be
d \vert 0_f(t) \dket = - i \hat{H}_{f,t}^U dt \circ \vert 0_f(t) \dket,
\label{unitary stoch Liouville Strat}
\ee
of the Stratonovich type is specified with the stochastic hat-Hamiltonian
\be
\hat{H}_{f,t}^U dt = \hat{H}_S dt + d \hat{M}_t^U,
\ee
with the Hermitian martingale operator
\be
\left( d\hat{M}_t^U \right)^\dagger = d\hat{M}_t^U.
\label{condition for unitary martingale}
\ee
Note that (\ref{unitary stoch Liouville Strat}) does not 
satisfy generally the conservation of probability 
just within the relevant system, i.e.,
\be
\bra 1 \vert d \hat{M}_t^U \neq 0,
\ee
but it does within whole the system, the relevant and irrelevant
systems, i.e.,
\be
\dbra 1 \vert d \hat{M}_t^U = 0.
\ee
Here, $\dbra 1 \vert = \bra 1 \vert \bra \vert$ with $\bra \vert$ 
being the bra-vacuum of the quantum Brownian motion (see 
Appendix~\ref{q Brownian motion}).
The martingale operator satisfies the fluctuation-dissipation 
theorem 
\be
d \hat{M}_t^U d \hat{M}_t^U = -2 \hat{\itPi} dt,
\label{f-d th for unitary}
\ee
of the second kind.

Introducing the unitary stochastic time-evolution operator 
$\hat{U}_f(t)$ by
\be
d \hat{U}_f(t) = - i \hat{H}_{f,t}^U dt \circ
\hat{U}_f(t),
\ee
with the initial condition $\hat{U}_f(0) = 1$,
we can define the stochastic Heisenberg operator
\be
A(t) = \hat{U}_f^{-1}(t) A \hat{U}_f(t),
\label{stoch Heisenberg op}
\ee
which satisfies the stochastic Heisenberg equation
(the Langevin equation)
\be
dA(t) = i [ \hat{H}_f^U(t) dt \strat \ A(t) ],
\ee
of the Stratonovich type.
Note that the time-evolution generator $\hat{U}_f(t)$ 
is a unitary operator:
\be
\hat{U}_f^\dagger(t) = \hat{U}_f^{-1}(t).
\ee

By making use of the relation between the Ito and Stratonovich
stochastic calculus, we can derive from 
(\ref{unitary stoch Liouville Strat}) 
the stochastic Liouville equation 
\be
d \vert 0_f(t) \dket = - i \hat{\cal H}_{f,t}^U dt \vert 0_f(t) \dket,
\label{unitary stoch Liouville Ito}
\ee
of the Ito type with the stochastic hat-Hamiltonian
\be
\hat{\cal H}_{f,t}^U = \hat{H} dt + d \hat{M}_t^U.
\ee
Applying $\bra \vert$ to (\ref{unitary stoch Liouville Ito}), 
we see easily that it reduces to 
the quantum master equation (\ref{master eq}).

Within the Ito calculus, the time-evolution operator
$\hat{U}_f(t)$ satisfies
\be
d \hat{U}_f(t) = - i \hat{{\cal H}}_{f,t}^U dt
\hat{U}_f(t),
\ee
with the initial condition $\hat{U}_f(0)=1$.
The stochastic Heisenberg operator $A(t)$ defined 
by (\ref{stoch Heisenberg op})
satisfies the stochastic Heisenberg equation
\bea
dA(t) \aeq i [ \hat{\cal H}_f^U(t) dt,\ A(t) ] 
\nonumber\\
&&- d\hat{M}^U(t)\  [ d\hat{M}^U(t),\ A(t) ],
\eea
of the Ito type.
Here, we introduced the martingale operator in the 
Heisenberg representation by
\be
d \hat{M}^U(t) = d \left( \hat{U}_f^{-1}(t) M_t^U \hat{U}_f(t) \right).
\ee
Note that \cite{cloud chamber}
\be
d \hat{M}^U(t) = d' \hat{M}^U(t),
\ee
with
\be
d' \hat{M}^U(t) =  \hat{U}_f^{-1}(t) d M_t^U \hat{U}_f(t).
\ee

\section{Application to Quantum Damped Harmonic Oscillator}
\label{damped oscillator system}

\subsection{Quantum Master Equation}

The hat-Hamiltonian of the semi-free field is bi-linear in $(a, \tilde{a},
a^\dagger, \tilde{a}^\dagger)$, and is
invariant under the phase transformation
$a \rightarrow a \me^{i \theta}$:
\be
\hat{H} =  g_1 a^\dagger a + g_2 \tilde{a}^\dagger \tilde{a}
	+ g_3 a \tilde{a} + g_4 a^\dagger \tilde{a}^\dagger
	+ g_0,
\label{hat-H0}
\ee
where $g$'s are time-dependent c-number complex 
functions.

The operators \( a,\ \tilde{a}^\dagger \), etc.\ satisfy the canonical 
commutation relation:\footnote{
Throughout this paper, we confine ourselves to the 
case of boson fields, for simplicity. The extension to
the case of fermion fields are rather straightforward.
}
\be
 [ a_{\bf k},\ a_{\bf k'}^\dagger ] = \delta_{\bf k,k'},
 \qquad 
 [ \tilde{a}_{\bf k},\ \tilde{a}_{\bf k'}^\dagger ] = 
 \delta_{\bf k,k'}.
\label{canonical commutation}
\ee
The tilde and non-tilde operators are mutually commutative.
Throughout this paper, we do not label explicitly  
the operators \( a,\ \tilde{a}^\dagger \), etc.\ 
with a subscript \( {\bf k} \)
for specifying a 
momentum and/or other degrees of freedom.  However, remember 
that we are dealing with a {\em dissipative quantum field}.

The tildian nature $(i \hat{H})^\sim = i \hat{H}$ makes 
(\ref{hat-H0}) tildian:
\be
\hat{H} = \omega (a^\dagger a - \tilde{a}^\dagger \tilde{a})
	+ i \hat{\itPi},
	\label{H-t}
\ee
with
\be
\hat{\itPi} = c_1 (a^\dagger a + \tilde{a}^\dagger \tilde{a})
	+ c_2 a \tilde{a} + c_3 a^\dagger \tilde{a}^\dagger
	+ c_4,
\label{pi-1}
\ee
where $\omega= \Re\mbox{e}\ g_1= - \Re\mbox{e}\ g_2,\ 
c_1 = \Im\mbox{m}\ g_1=\Im\mbox{m}\ g_2,\ c_2=\Im\mbox{m}
\ g_3,\ c_3= \Im\mbox{m}\ g_4$ and $c_4= \Im\mbox{m}\ g_0$.

With the help of (\ref{thermal state cond}) for $A=a$:
\be
\bra 1 \vert \tilde{a} = \bra 1 \vert a^\dagger,
\label{initial bra}
\ee
the property $\bra 1 \vert \hat{H} =0$ gives us the relations
\be
2 c_1 + c_2 + c_3 = 0,\qquad 
c_3 + c_4 = 0.
\ee
Then, (\ref{pi-1}) reduces to
\bea
\hat{\itPi} \aeq c_1 (a^\dagger a + \tilde{a}^\dagger \tilde{a})
\nonumber\\
&& - c_2 a \tilde{a} - \left( 2 c_1 + c_2 \right) 
	a^\dagger \tilde{a}^\dagger + \left( 2 c_1 + c_2 \right).
\label{pi-2}
\eea

Let us write down here the Heisenberg equations for $a$ and 
$a^\dagger$:
\bea
\frac{d}{dt} a(t) \aeq -i\omega a(t) + c_1 a(t)  
-\left(2 c_1 + c_2 \right) \tilde{a}^{\dd}(t), 
\label{H eq for a}\\
\frac{d}{dt} a^{\dd}(t) \aeq i\omega a^{\dd}(t) - c_1 a^{\dd}(t)  
-c_2 \tilde{a}(t).
\label{H eq for a-d}
\eea
Since the semi-free hat-Hamiltonian $\hat{H}$
is not necessarily Hermite, we introduced the symbol $\dd$
in order to distinguish it from the Hermite conjugation 
$\dagger$.
However in the following, we will use $\dagger$ instead of $\dd$, 
for simplicity, unless it is confusing.
By making use of the Heisenberg equations (\ref{H eq for a}) and 
(\ref{H eq for a-d}), we obtain the equation of motion for
a vector $\bra 1\vert a^\dagger(t) a(t)$ in the form
\be
\frac{d}{dt} \bra 1 \vert a^\dagger(t) a(t) = - 2\kappa 
\bra 1 \vert a^\dagger(t) a(t) + i \itSigma^< \bra 1 \vert,
\label{eq for a vector}
\ee
where we introduced $\kappa$ 
and $\itSigma^<$ respectively by
\bea
\kappa \aeq c_1 + c_2,
\label{kappa}\\
\itSigma^< \aeq i(2 c_1 + c_2).
\label{sigma<}
\eea
In deriving (\ref{eq for a vector}), we used the thermal state condition
(\ref{thermal state cond}) in order to eliminate tilde operators.

Applying the thermal ket vacuum $\vert 0 \ket$ 
to (\ref{eq for a vector}), 
we obtain the equation of motion for the {\it one-particle 
distribution function}
\begin{equation}
 n(t) =\  \bra1\vert a^{\dd}(t) a(t) \vert 0\ket
= \bra 1 \vert a^{\dagger} a \vert 0(t) \ket,
\label{one-p1}
\end{equation}
as
\be
\frac{d}{dt} n(t) = -2 \kappa n(t) +i \itSigma^<.
\label{a2}
\ee
The equation (\ref{a2}) is the Boltzmann equation of the
system.
The function $\itSigma^<$ is given when the
interaction hat-Hamiltonian is specified.

The initial ket-vacuum $\vert 0 \ket =\vert 0(t=0) \ket$
is specified by
\be
a \vert 0 \ket = f \tilde{a}^\dagger \vert 0 \ket,
\label{initial ket}
\ee
with a real quantity $f$. 
Here, we are neglecting the {\it initial correlation} 
\cite{Fujita}.
The initial condition of the one-particle distribution 
function $n=n(t=0)$ can be derived by treating $\bra 1 \vert
 a \tilde{a} \vert 0 \ket$ as follows.
In the first place,
\bea
\bra 1 \vert a \tilde{a} \vert 0 \ket \aeq \bra 1 \vert a
f a^\dagger \vert 0 \ket \nonumber\\
\aeq f \left( \bra 1 \vert a^\dagger a \vert 0\ket + 
\bra 1 \vert 0 \ket \right) \nonumber\\
\aeq f \left( n + 1 \right),
\label{n-f 1}
\eea
where we used the tilde conjugate of (\ref{initial ket}) 
for the first equality, and the canonical commutation relation 
(\ref{canonical commutation}) for the second.
On the other hand,
\bea
\bra 1 \vert a \tilde{a} \vert 0 \ket \aeq \bra 1 \vert \tilde{a} a 
\vert 0 \ket \nonumber\\
\aeq \bra 1 \vert a^\dagger a \vert 0 \ket \nonumber\\
\aeq n.
\label{n-f 2}
\eea
Here, for the first equality, we used (\ref{commutable}),
i.e., the commutativity between the tilde and non-tilde operators,
and, for the second equality, (\ref{initial bra}).
Equating (\ref{n-f 1}) and (\ref{n-f 2}), we see that 
\be
n = \frac{f}{1-f},\qquad \left(f = \frac{n}{1+n} \right).
\label{n-f}
\ee

If it is assumed that there is only one stationary state,
we can refer the stationary state as a thermal equilibrium state.
We will assign the thermal equilibrium state to be 
specified by the Planck distribution function with temperature $T$:
\be
n(t\rightarrow \infty) = \bar{n} = \frac{1}{\me^{\omega/T} -1}.
\ee
Then, we have from (\ref{a2})
\be
i \itSigma^< = 2 \kappa \bar{n}.
\label{Sigma}
\ee
In this case, the Boltzmann equation (\ref{a2}) reduces to
\be
\frac{d}{dt} n(t) = -2 \kappa \left( n(t) - \bar{n} \right).
\label{Boltzmann eq for stationary process}
\ee

Solving (\ref{kappa}) and (\ref{sigma<}) with respect to $c_1$ and
$c_2$, and substituting (\ref{Sigma}) for $\itSigma^<$ 
into (\ref{pi-1}), we 
finally arrive at
the most general form of the semi-free hat-Hamiltonian 
$\hat{H}$ corresponding to the stationary process 
\cite{netfd3}:
\be
 \hat{H} = \hat{H}_S + i\hat{\itPi},
\label{b2}
\ee
where
\be
\hat{H}_S = H_S -\tilde{H}_S,\qquad 
H_S = \omega a^\dagger a,
\label{H-hat S}
\ee
\bea
\hat{\itPi} \aeq  - \kappa \left[ \left(1+2\bar{n}\right) \left( a^\dagger 
 a + \tilde{a}^\dagger \tilde{a} \right) 
\right. \nonumber\\
&& \left. - 2\left(1+\bar{n}\right)
 a \tilde{a} - 2\bar{n} a^\dagger \tilde{a}^\dagger \right] 
 - 2\kappa\bar{n}.
\label{Pi hat}
\eea
This hat-Hamiltonian is the same expression as 
the one derived by means of the 
principle of correspondence \cite{netfd1,netfd2} when NETFD was constructed
first by referring to the projection operator formalism of the damping theory 
\cite{Haake,tcl,general tcl}.  
The hat-Hamiltonian (\ref{b2}) with (\ref{H-hat S}) and (\ref{Pi hat}) describes 
time-evolution of the system of a damped harmonic oscillator.

A detailed investigation of the system is given in Appendix~\ref{new aspect}.
Here we just write down an attractive expression which leads us to 
a new concept, named {\it spontaneous creation of dissipation}. 
The expression is given by
\be
\vert 0(t) \ket = \exp \left[ - \int d^3k \bra 1 \vert \gamma_{k,t} 
\tilde{\gamma}_{k,t} \vert 0 \ket \gamma_k^{\venus} \tilde{\gamma}_k^{\venus}
\right] \vert 0 \ket
\ee
where 
\be
\bra 1 \vert \gamma_{k,t} \tilde{\gamma}_{k,t} \vert 0 \ket
= -n_k(t) + n_k(0)
\ee
is the {\it order parameter} for dissipative time-evolution of 
the unstable vacuum.
The annihilation and creation operators 
\bea
\gamma_{k,t}^{\mu=1} = \gamma_{k,t},\qquad && 
\gamma_{k,t}^{\mu=2} = \tilde{\gamma}_k^{\venus},
\\
\bar{\gamma}_{k,t}^{\mu=1} = \gamma_k^{\venus},\qquad  &&
\bar{\gamma}_{k,t}^{\mu=2} = - \tilde{\gamma}_{k,t},
\eea
are defined by
\be
 \gamma_{k,t}^\mu = B_k(t)^{\mu\nu} a_k^\nu,\qquad 
\bar{\gamma}_{k,t}^\mu = \bar{a}_k^\nu B_k^{-1}(t)^{\nu\mu},
\label{gamma Schroedinger}
\ee
with the {\em time-dependent Bogoliubov transformation}:
\begin{equation}
 B_k(t)^{\mu\nu} = \left(
 \begin{array}{cc}
  1+n_k(t) & -n_k(t) \\
  -1 & 1 \\
 \end{array}
 \right).
\label{time-dep B}
\end{equation}
They satisfy the canonical commutation relation
\be
[\gamma_{k,t}^\mu,\ \bar{\gamma}_{k',t}^\nu] = \delta_{k,k'} \delta^{\mu\nu},
\ee
and annihilate the bra- and ket-vacuums at time $t$:
\begin{equation}
 \gamma_{k,t} \vert 0(t) \ket \ = 0,
 \qquad \bra1\vert \tilde{\gamma}_k^{\venus} = 0.
\label{annihilate vacuum at t}
\end{equation}

\subsection{Non-Unitary Time-Evolution}

Confining ourselves to the case where
the interaction hat-Hamiltonian between the relevant system and 
the irrelevant system of Brownian motion
is bi-linear in ($a,\ a^\dagger,$ and their tilde conjugates) and
($dB_t,\ dB^\dagger_t,$ and their tilde conjugates), and is invariant 
under the phase transformation
$a \rightarrow a \me^{i \theta}$, and $dB_t \rightarrow dB_t\ 
\me^{i \theta}$,
the martingale operator satisfying 
(\ref{condition for non-unitary martingale}) is given by
\be
:d\hat{M}_t:\ = i \left( \gamma^{\venus} dW_t
	+  \tilde{\gamma}^{\venus} d\tilde{W}_t \right),
	\label{martingale with lambda}
\ee
where we introduced the random force operator
\be
dW_t = \sqrt{2\kappa} \left( \mu dB_t + 
\nu d\tilde{B}^\dagger_t \right),
\label{W-F0} \qquad
\ee
and the annihilation and creation operator 
\be
\gamma_\nu = \mu a + \nu \tilde{a}^\dagger, \qquad
\gamma^{\venus} = a^\dagger - \tilde{a},
\ee
of the relevant system with $\mu + \nu =1$, 
which satisfy the commutation relation 
\be
[\gamma_\nu,\ \gamma^{\venus}] = 1,
\ee
and annihilate the relevant bra-vacuum:
\be
\bra 1 \vert \gamma^{\venus} = 0,\qquad 
\bra 1 \vert \tilde{\gamma}^{\venus} = 0.
\ee
Note that the normal ordering $:\cdots:$ in (\ref{martingale with lambda}) 
is defined with respect to
the annihilation and the creation operators.
Making use of the annihilation and the creation operators,
we can rewrite $\hat{\itPi}_R$ and $\hat{\itPi}_D$ consisting of 
$\hat{\itPi}$ introduced in (\ref{Pi hat}) as
\bea
\hat{\itPi}_R \aeq - \kappa
	\left( \gamma^{\venus} \gamma_\nu
	+ \tilde{\gamma}^{\venus} \tilde{\gamma}_\nu \right), 
\label{Pi_R}
\\
\hat{\itPi}_D \aeq 2 \kappa \left( \bar{n} + \nu \right) 
	\gamma^{\venus} \tilde{\gamma}^{\venus}.
\label{Pi_D}
\eea

The Langevin equations for $a(t)$ and $a^\dagger(t)$ are
given by
\bea
da(t) \aeq -i \omega a(t) dt  
\nonumber\\
&&- \kappa 
[ ( \mu - \nu ) a(t) + 2 \nu 
	\tilde{a}^\dagger(t) ] dt + dW_t, 
\label{lan-a-1}\\
da^\dagger(t) \aeq i \omega a^\dagger(t) dt 
\nonumber\\
&&	- \kappa [ 2 \mu \tilde{a}(t) - ( \mu - \nu ) 
a^\dagger(t) ] dt + d\tilde{W}_t,
\label{lan-a-2}
\eea
where we used the facts
\be
dW(t) = dW_t,
\qquad
d\tilde{W}(t) = d\tilde{W}_t.
\ee
If we put $\nu=1/2$, then $\mu=1/2$, (\ref{lan-a-1}) and 
(\ref{lan-a-2}) reduce, respectively, to
\bea
da(t) \aeq -i \omega a(t) dt - \kappa 
	\tilde{a}^\dagger(t)  dt + dW_t, 
\label{lan-a-1a}\\
da^\dagger(t) \aeq i \omega a^\dagger(t) dt
	- \kappa \tilde{a}(t) dt + d\tilde{W}_t.
\label{lan-a-2a}
\eea
Although $dW_t$ and $d\tilde{W}_t$ are commutative, 
we have the conservation of 
the equal-time canonical commutation relation
\be
d \left( [a(t),\ a^\dagger(t) ] \right) =0.
\ee

Applying the bra-vacuum $\dbra 1 \vert$ 
to (\ref{lan-a-1}) and (\ref{lan-a-2}), 
we have, for any value of $\nu$,
\bea
d \dbra 1 \vert a(t) \aeq -i \omega \dbra 1 \vert a(t) dt 
\nonumber\\
&&- \kappa 
\dbra 1 \vert a(t)  dt + \sqrt{2\kappa} \dbra 1 \vert d B_t, 
\label{lan-a-1aa}\\
d\dbra 1 \vert a^\dagger(t) \aeq i \omega \dbra 1 \vert a^\dagger(t) dt 
\nonumber\\
&&	- \kappa \dbra 1 \vert a^\dagger(t) dt 
+ \sqrt{2\kappa} \dbra 1 \vert d B_t^\dagger.
\label{lan-a-2aa}
\eea
Note that these Langevin equations for the vector $\dbra 1 \vert a(t)$ 
and $\dbra 1 \vert a^\dagger(t)$ have, respectively, 
the same structure as (\ref{Lan a}) and (\ref{Lan a dag}).

\subsection{Unitary Time-Evolution}

The unitary martingale operator satisfying 
(\ref{condition for unitary martingale}) is given by
\bea
:d\hat{M}_t^U:\ \aeq i \sqrt{2\kappa} : \left( a^\dagger dB_t 
- dB^\dagger_t a + \mbox{t.c.} \right):
\nonumber\\
\aeq i \left( \gamma^{\venus} dW_t
	+  \tilde{\gamma}^{\venus} d\tilde{W}_t \right)
\nonumber\\
&& - i \left( dW^{\venus}_t \gamma_\nu
	+  d\tilde{W}^{\venus}_t \tilde{\gamma}_\nu \right),
	\label{unitary martingale}
\eea
with t.c indicating tilde conjugate.
Note that there is no cross term between tilde and non-tilde 
operators.
Here, we introduced new random force operators 
\be
dW^{\venus}_t = \sqrt{2\kappa} \left(dB^\dagger_t - 
d\tilde{B}_t \right),
\ee
which annihilates the bra-vacuum $\bra \vert$ of the
irrelevant system:
\be
\bra \vert dW^{\venus}_t= 0,\qquad \bra \vert 
d\tilde{W}^{\venus}_t= 0,
\ee
and satisfies the commutation relation
\be
[d W_t,\ d W_t^{\venus} ] = 2\kappa.
\ee
The expression (\ref{unitary martingale}) is consistent with 
the microscopic Hamiltonian of the linear dissipative coupling.

The martingale operator (\ref{unitary martingale}) satisfies 
the fluctuation-dissipation theorem (\ref{f-d th for unitary})
with (\ref{Pi hat}).
Therefore, we conclude that there exists the system of 
stochastic differential equations with unitary time-evolution 
operator consistent with the quantum master equation 
(\ref{master eq}) with the hat-Hamiltonian (\ref{b2}).

The Langevin equations for $a(t)$ and $a^\dagger(t)$ are given by
\bea
da(t) \aeq -i \omega a(t) dt + \sqrt{2\kappa} \  d B(t),
\label{Lan eq a1 unitary}
\\
da^\dagger(t) \aeq i \omega a^\dagger(t) dt 
+ \sqrt{2\kappa} \  d B^\dagger(t),
\label{Lan eq a2 unitary}
\eea
where the operators $dB(t)$ and $dB^\dagger(t)$ in 
the Heisenberg representation are defined by
\bea
\sqrt{2\kappa} \  dB(t) \aeq \hat{U}_f^{-1}(t) \circ 
\sqrt{2\kappa} \  dB_t \circ \hat{U}_f(t)
\nonumber\\
\aeq \sqrt{2\kappa} \  dB_t - \kappa a(t) dt,
	\label{Heisenberg B-1}
\\
\sqrt{2\kappa} \  dB^\dagger(t) \aeq \hat{U}_f^{-1}(t) \circ 
\sqrt{2\kappa} \  dB_t^\dagger \circ \hat{U}_f(t)
\nonumber\\
\aeq \sqrt{2\kappa} \  dB_t^\dagger - \kappa a^\dagger(t) dt.
	\label{Heisenberg B-2}
\eea
In deriving (\ref{Heisenberg B-1}) and (\ref{Heisenberg B-2}), 
we used the properties
\bea
[\hat{U}_f(t) \strat \ \sqrt{2\kappa} \  dB_t ] 
\aeq [ \hat{U}_f(t),\ \sqrt{2\kappa} \  dB_t ]
\nonumber\\
&& + \frac12 [ d\hat{U}_f(t),\ \sqrt{2\kappa} \  dB_t ]
\nonumber\\
\aeq \kappa a \hat{U}_f(t) dt,
\label{comm:S-W-1}
\eea
\be
[ \hat{U}_f(t) \strat \ \sqrt{2\kappa} \  dB_t^\dagger ]
= \kappa a^\dagger \hat{U}_f(t) dt.
\label{comm:S-W-2}
\ee
and
\be
[ \hat{U}_f(t),\ dB_t ] 
	= [ \hat{U}_f(t),\ dB_t^\dagger ]
	= 0,
\ee
which comes from the characteristics of the Ito multiplication:
\be
\bra  \vert \hat{U}_f(t) dB_t \vert \ket
	= \bra \vert \hat{U}_f(t) dB_t^\dagger \vert \ket
	= 0.
\ee

Substituting (\ref{Heisenberg B-1}) and (\ref{Heisenberg B-2}), 
and applying $\dbra 1 \vert$, 
(\ref{Lan eq a1 unitary}) and (\ref{Lan eq a2 unitary}) 
reduce to (\ref{lan-a-1aa}) and (\ref{lan-a-2aa})
having the same structures as (\ref{Lan a}) 
and (\ref{Lan a dag}), respectively.

\section{Application to Quantum Kramers Equation}
\label{Kramers system}

\subsection{Quantum Master Equation}

Let us find out the general structure of hat-Hamiltonian 
which is bilinear in $(x,\ p,\ \tilde{x},\ \tilde{p})$.
$x$ and $p$ satisfies the canonical commutation relation 
\be
[x,\ p] = i.
\ee
Accordingly, $\tilde{x}$ and $\tilde{p}$ satisfies 
\be
[\tilde{x},\ \tilde{p}] = -i.
\ee

The conditions, $(i \hat{H})^\sim = i \hat{H}$, and
$\bra 1 \vert \hat{H} = 0$ give us the general expression
\be
\hat{H} = \hat{H}_S + i \hat{\itPi},
\ee
where
\be
\hat{H}_S = H_S - \tilde{H}_S, \quad 
H_S = \frac{1}{2m} p^2 + \frac{m \omega^2}{2} x^2,
\ee
\be
\hat{\itPi} = \hat{\itPi}_R + \hat{\itPi}_D,
\ee
with
\bea
\hat{\itPi}_R \aeq - i \frac{1}{2} \kappa \left( x - \tilde{x} \right)
\left( p + \tilde{p} \right),
\nonumber\\
\hat{\itPi}_D \aeq - \frac{1}{2} \kappa m \omega (1+2\bar{n})
\left( x - \tilde{x} \right)^2.
\eea
Here, we neglected the diffusion in $x$-space.
The Schr\"odinger equation 
\be
\frac{\partial}{\partial t} \vert 0(t) \ket = -i \hat{H}
\vert 0(t) \ket,
\label{q Kramers eq}
\ee
gives the {\it quantum Kramers equation} \cite{Arimitsu1982-2}.

The Heisenberg equation for the dissipative system is 
given by
\bea
\frac{d}{dt}x(t) \aeq i [ \hat{H}(t),\ x(t) ]
\nonumber\\
\aeq \frac{1}{m} p(t)+ \frac12 
\kappa \left\{ x(t) - \tilde{x}(t) \right\},
\label{H eq x}
\\
\frac{d}{dt}p(t) \aeq - m\omega^2 x(t) 
- \frac{1}{2} \kappa \left\{ p(t) + \tilde{p}(t) \right\}
\nonumber\\
&&+ i \kappa m \omega 
(1+2\bar{n}) \left\{ x(t) - \tilde{x}(t) \right\}.
\label{H eq p}
\eea

Applying the bra-vacuum $\bra 1 \vert$ of the relevant system,
we have the equations for the vectors:
\bea
\frac{d}{dt}\bra 1 \vert x(t) \aeq \frac{1}{m} \bra 1 \vert p(t),
\nonumber\\
\frac{d}{dt} \bra 1 \vert p(t) \aeq - m\omega^2 \bra 1 \vert x(t) 
- \kappa \bra 1 \vert p(t).
\eea

\subsection{Non-Unitary Time-Evolution}

The stochastic Liouville equation within the Ito calculus becomes
\be
d \vert 0_f(t) \dket = -i \hat{{\cal H}}_{f,t} dt 
\vert 0_f(t) \dket,
\label{q stoch Kramers eq}
\ee
with the stochastic hat-Hamiltonian 
\be
\hat{{\cal H}}_{f,t} dt = \hat{H} dt + d\hat{M}_t.
\ee
Here, the martingale operator $d\hat{M}_t$ satisfying 
(\ref{condition for non-unitary martingale}) is defined by
\be
d\hat{M}_t = \left( x - \tilde{x} \right) \left(
dX_t + d\tilde{X}_t \right),
\ee
with 
\be
dX_t = \frac{\sqrt{\kappa m \omega}}{2} \left( dB_t + dB_t^\dagger
\right),
\ee
where $dB_t$, $dB_t^\dagger$ and their tilde conjugates are 
the operators representing quantum Brownian motion (see
Appendix~\ref{q Brownian motion}).
The generalized fluctuation-dissipation theorem is given by
\be
d\hat{M}_t d\hat{M}_t = -2\hat{\itPi}_D dt.
\ee
Taking a random average, the stochastic Liouville equation 
(\ref{q stoch Kramers eq}) reduces to 
the quantum master equation (\ref{q Kramers eq})
with 
$
\vert 0(t) \ket = \bra \vert 0_f(t) \dket.
$

The stochastic Heisenberg equation (the Langevin equation)
for this hat-Hamiltonian is given by
\bea
d x(t) \aeq i [ \hat{{\cal H}}_{f}(t) dt,\ x(t) ]
- d\hat{M}(t) \ [d\hat{M}(t),\ x(t) ]
\nonumber\\
\aeq \frac{1}{m} p(t) dt + \frac12 
\kappa \left\{ x(t) - \tilde{x}(t) \right\} dt,
\label{stoch H eq x}
\\
d p(t) \aeq - m\omega^2 x(t) dt
- \frac{1}{2} \kappa \left\{ p(t) + \tilde{p}(t) \right\} dt
\nonumber\\
&& - \left( dX_t + d \tilde{X}_t \right),
\label{stoch H eq p}
\eea
where we used the properties
\be
dX(t) = dX_t,\qquad 
d\tilde{X}(t) = d\tilde{X}_t.
\ee

Applying the bra vacuum $\dbra 1 \vert$ to (\ref{stoch H eq x}) and 
(\ref{stoch H eq p}), we have the Langevin equations for 
vectors
\bea
d \dbra 1 \vert x(t) \aeq \frac{1}{m} \dbra 1 \vert p(t) dt,
\label{stoch H eq x vec}
\\
d \dbra 1 \vert p(t) \aeq - m\omega^2 \dbra 1 \vert x(t) dt
- \kappa \dbra 1 \vert p(t) dt
\nonumber\\
&& - 2 \dbra 1 \vert dX_t.
\label{stoch H eq p vec}
\eea

The averaged equation of motion is given by applying 
$\vert 0 \ket$ to (\ref{stoch H eq x vec}) and 
(\ref{stoch H eq p vec}) in the forms
\bea
\frac{d}{dt} \dbra x(t) \dket \aeq \frac{1}{m} \dbra p(t) \dket,\\
\frac{d}{dt} \dbra p(t) \dket \aeq - m\omega^2 \dbra x(t) \dket
- \kappa \dbra p(t) \dket,
\eea
where $\dbra \cdots \dket = \bra 1 \vert \bra \vert \cdots \vert \ket 
\vert 1 \ket$. The vacuums $\bra \vert$ and $\vert \ket$ are introduced in 
Appendix \ref{q Brownian motion}.
These averaged equations can be also derived from (\ref{H eq x})
and (\ref{H eq p}) by taking the average $\dbra \cdots \dket$.

\subsection{Unitary Time-Evolution}

The martingale operator representing position-position
interaction may be given by
\be
dM_t^U = x dX_t - \tilde{x} d\tilde{X}_t.
\ee
We did not include the crossing terms between 
tilde and non-tilde operators to be consistent with 
the microscopic interaction Hamiltonian.

The fluctuation-dissipation theorem for this martingale
operator is given by
\be
dM_t^U dM_t^U = -2 \hat{\itPi}^U dt,
\ee
with 
\be
\hat{\itPi}^U = - \frac{\kappa m \omega}{8} (1+2\bar{n}) 
\left( x - \tilde{x} \right)^2.
\ee

Then, the Ito stochastic hat-Hamiltonian becomes
\be
\hat{\cal H}_{f,t}^U dt = \hat{H}^U dt + dM_t^U,
\ee
where 
\be
\hat{H}^U = \hat{H}_S + i \hat{\itPi}^U.
\ee
is the hat-Hamiltonian for the master equation.
The master equation is different from (\ref{master eq}).

The stochastic Heisenberg equations (the Langevin equations)
for $x(t)$ and $p(t)$ become
\bea
dx(t) \aeq \frac{1}{m} p(t) dt,
\label{Lan x Kramers unitary}
\\
dp(t) \aeq - m \omega^2 x(t) dt - dX_t,
\label{Lan p Kramers unitary}
\eea
where we used the fact 
\be
dX(t) = dX_t.
\ee

Applying $\dbra 1 \vert$ to (\ref{Lan x Kramers unitary}) 
and (\ref{Lan p Kramers unitary}), we have the Langevin equations
for the vectors $\dbra 1 \vert x(t)$ and $\dbra 1 \vert p(t)$ 
in the forms
\bea
d \dbra 1 \vert x(t) \aeq \frac{1}{m} \dbra 1 \vert p(t) dt,
\label{Lan vec x Kramers unitary}
\\
d \dbra 1 \vert p(t) \aeq -m \omega^2 \dbra 1 \vert x(t) dt
- \dbra 1 \vert X_t,
\eea
which are different from (\ref{stoch H eq x vec}) 
and (\ref{stoch H eq p vec}).

\section{Summary and Discussion}
\label{summary}

Within the system of non-unitary time-evolution generator
(non-unitary system),
the time-evolution generator $V_f(t)$ is constituted by 
the commutative random force operators $dW_t$ and $d\tilde{W}_t$.
Therefore, the random force operators $dW(t)$ and $dX(t)$ in 
the Heisenberg representation is, respectively, equal to $dW_t$ and 
$dX_t$ in the Schr\"odinger representation, i.e.,
\be
dW(t) = dW_t,\qquad dX(t) = dX_t.
\ee

In the application of the system of unitary time-evolution generator
(unitary system) to the damped harmonic oscillator 
where the martingale operator 
is constituted by non-commutative random force operators
manifesting the linear dissipative coupling between the relevant and
irrelevant sub-systems, the random force operators in 
the Heisenberg representation are related to those in 
the Schr\"odinger representation by 
\bea
dW(t) \aeq dW_t - \kappa \gamma_\nu(t) dt,
\\
dW^{\venus}(t) \aeq dW_t^{\venus} - \kappa \gamma^{\venus}(t) dt.
\eea
The second terms show up because of the non-commutativity.
The appearance of these terms is essential in order to make 
the unitary system consistent with corresponding master equation.

On the contrary, in the application of the unitary system 
to the quantum Kramers equation where
the martingale operator is constituted only by commutative 
random force operators manifesting the position-position 
coupling between the relevant and irrelevant sub-systems,
the random force operators in the Heisenberg representation 
is equal to those in the Schr\"odinger representation, i.e.,
\be
dX(t) = dX_t.
\ee
Therefore, the unitary system cannot be consistent with 
corresponding master equation.

The above applications tells us that the origin of dissipation 
cannot be quantum mechanical. 
In spite of this unsatisfactory nature of the unitary system,
it is attractive since hat-Hamiltonian for microscopic system
is Hermitian and there is no mixing terms between 
tilde and non-tilde operators. The hat-Hamiltonian should have
the structure 
\be
\hat{H} = H - \tilde{H},\qquad H^\dagger = H,
\ee
for microscopic systems.
In fact, we succeeded to extract the correct stochastic hat-Hamiltonian
for the stochastic Kramers equation by an appropriate coarse graining
of operators (the {\it stochastic mapping}) in time and corresponding
renormalization of physical quantities \cite{Kramers eq}.
The simple limit \cite{Accardi-1995} does not 
give us the correct Kramers equation.
This something touchy situation should be investigated 
based on the unified system of stochastic differential equations
shown in this paper. 
It will be reported in the future publications.

\acknowledgements

The authors would like to thank Dr.~N.~Arimitsu, 
Dr.~T.~Saito, Dr.~A.~Tanaka, Mr.~T.~Imagire, Mr.~T.~Indei 
and Mr.~Y.~Endo for their collaboration with fruitful discussions.

\appendix

\section{New Aspect for A Damped Harmonic Oscillator}
\label{new aspect}

\subsection{Thermal Doublet}

Let us introducing the thermal doublet notation by
\bea
a(t)^{\mu = 1} = a(t),\quad &&
a(t)^{\mu = 2} = \tilde{a}^{\dagger}(t),
\\
\bar{a}(t)^{\mu = 1} = a^{\dagger}(t),\quad &&
\bar{a}(t)^{\mu = 2} = - \tilde{a}(t).
\eea
Then, the canonical commutation relation can be written as
\be
[a(t)^\mu,\ \bar{a}(t)^\nu ] = \delta^{\mu\nu}.
\ee
Note that 
\be
a(t)^\mu = \hat{V}^{-1}(t) a^\mu \hat{V}(t),\quad
\bar{a}(t)^\mu = \hat{V}^{-1}(t) \bar{a}^\mu \hat{V}(t).
\ee

Making use of the thermal doublet notation, the hat-Hamiltonian
(\ref{b2}) reduces to 
\bea
\hat{H} \aeq \omega \bar{a}^\mu a^\mu + i \hat{\itPi} + \omega,
\label{hat-H doublet}
\\
\hat{\itPi} \aeq -\kappa \bar{a}^\mu A^{\mu\nu} a^\nu + \kappa,
\eea
with
\be
 A^{\mu\nu} = \left(
 \begin{array}{cc}
  1+ 2\bar{n} & -2\bar{n} \\
  2(1+\bar{n}) & -(1 + 2 \bar{n})\\
 \end{array}
 \right).
\label{A}
\ee

The Heisenberg equations for the semi-free particle become
\bea
\frac{d}{dt} a(t)^\mu \aeq i [\hat{H}(t),\ a(t)^\mu]
\nonumber\\
\aeq -i \left[ \omega \delta^{\mu\nu} - i\kappa A^{\mu\nu} 
\right] a(t)^\nu,
\nonumber\\
\frac{d}{dt} \bar{a}(t)^\mu \aeq i [\hat{H}(t),\ \bar{a}(t)^\mu]
\nonumber\\
\aeq \bar{a}(t)^\nu i \left[ \omega \delta^{\nu\mu} - i\kappa A^{\nu\mu} 
\right].
\eea

\subsection{Annihilation and Creation Operators}

Let us introduce the annihilation and creation operators, 
\bea
\gamma(t)^{\mu = 1} = \gamma(t),\quad && 
\gamma(t)^{\mu = 2} = \tilde{\gamma}^{\venus}(t),
\\
\bar{\gamma}(t)^{\mu = 1} = \gamma^{\venus}(t),\quad &&
\bar{\gamma}(t)^{\mu = 2} = -\tilde{\gamma}(t),
\eea
by 
\begin{equation}
 \gamma(t)^\mu = B(t)^{\mu\nu} a(t)^\nu,\qquad \bar{\gamma}(t)^\mu = 
 \bar{a}(t)^\nu B^{-1}(t)^{\nu\mu},
\label{a4}
\end{equation}
with the {\em time-dependent Bogoliubov transformation}:
\begin{equation}
 B(t)^{\mu\nu} = \left(
 \begin{array}{cc}
  1+n(t) & -n(t) \\
  -1 & 1 \\
 \end{array}
 \right),
\label{a5}
\end{equation}
where $n(t)$ is the one-particle distribution function
satisfying the Boltzmann equation 
(\ref{Boltzmann eq for stationary process}).

The annihilation and creation operators satisfy
the canonical commutation relation
\be
[\gamma(t)^\mu,\ \bar{\gamma}(t)^\nu] = \delta^{\mu\nu},
\ee
and annihilate the bra- and ket-vacuums at the initial time:
\begin{equation}
 \gamma(t) \vert 0\ket \ = 0,\qquad \bra1\vert \tilde{\gamma}^{\venus}(t) = 0.
\label{a7}
\end{equation}

The equation of motion for the thermal doublet $\gamma(t)^\mu$ is 
derived as
\bea
\frac{d}{dt} \gamma(t)^\mu \aeq \frac{d B(t)^{\mu \nu}}{dt}
a(t)^\nu + B(t)^{\mu \nu} \frac{d}{dt} a(t)^\nu \nonumber\\
\aeq \left[\frac{dB(t)}{dt} B^{-1}(t) \right]^{\mu\nu}
\gamma(t)^{\nu} 
\nonumber\\
&& -i \left[B(t) \left( \omega \ 1 
- i \kappa A \right) B^{-1}(t) \right]^{\mu\nu} \gamma(t)^\nu
\nonumber\\
\aeq -i \left[ \omega \delta^{\mu \nu} - i \kappa 
\tau_3^{\mu \nu} \right] \gamma(t)^\nu,
\label{eq for gamma}
\eea
where the matrix \(\tau_3^{\mu \nu}\) is defined by
\be
\tau_3^{11} = - \tau_3^{22} = 1,\ \tau_3^{12} = \tau_3^{21} = 0.
\ee
For the third equality, we used the relations
\be
\frac{dB(t)}{dt} = \left(
 \begin{array}{cc}
  1 & -1 \\
  0 & 0 \\
 \end{array}
 \right)
\frac{dn(t)}{dt},
\ee
\be
\frac{dB(t)}{dt} B^{-1}(t) = 
- \frac{n(t)}{dt} \tau_+,
\ee
\be
B(t)  A  B^{-1}(t) = \tau_3 + 2\left[ n(t) - \bar{n}
\right] \tau_+,
\ee
where
\be
\tau_+ = \left(
 \begin{array}{cc}
  0 & 1 \\
  0 & 0 \\
 \end{array}
 \right).
\ee
The Boltzmann equation (\ref{Boltzmann eq for stationary process}) 
has been used also.

The solution of (\ref{eq for gamma}) is given by
\be
\gamma(t)^\mu = \exp \left\{-i\left(\omega
 \delta^{\mu \nu} -i\kappa \tau_3^{\mu \nu} \right) 
(t-t') \right\} \gamma(t')^\nu.
\label{solution of gamma}
\ee
This expression gives 
\bea
\bra 1 \vert \gamma(t) \gamma^{\venus}(t') \vert 0 \ket
\aeq \me^{-i(\omega -i\kappa)(t-t')}
\bra 1 \vert \gamma(t') \gamma^{\venus}(t') \vert 0 \ket
\nonumber\\
\aeq \me^{-i(\omega -i\kappa)(t-t')},
\label{gamma-gamma}
\\
\bra 1 \vert \tilde{\gamma}(t') \tilde{\gamma}^{\venus}(t) 
\vert 0 \ket 
\aeq \bra 1 \vert \gamma(t') \gamma^{\venus}(t) 
\vert 0 \ket^{\sim} 
\nonumber\\
\aeq \me^{-i(\omega +i\kappa)(t-t')}.
\label{gamma-gamma tilde}
\eea

\subsection{Two-Point Function (Propagator)}

The time-ordered two-point function \( G(t,t')^{\mu\nu} \) has the form
\begin{eqnarray}
 G(t,t')^{\mu\nu} \aeq -i \bra1\vert T \left[a(t)^\mu \bar{a}(t')^\nu 
 \right]  \vert0\ket  \nonumber\\
 \aeq \left[ B^{-1}(t) {\cal G}(t,t') B(t') \right]^{\mu\nu},
\label{two point func}
\end{eqnarray}
where
\begin{eqnarray}
 {\cal G}(t,t')^{\mu\nu} \aeq -i \bra1\vert T \left[ \gamma(t)^\mu \bar{\gamma}
 (t')^\nu \right] \vert0\ket  
 \nonumber\\
\aeq \left(
  \begin{array}{cc}
   G^R(t,t') & 0 \\
   0 & G^A(t,t') \\
  \end{array}
 \right),
\label{a6}
\end{eqnarray}
with
\begin{eqnarray}
 G^R(t,t') \aeq -i \theta(t-t') \me^{-i\left(\omega
 -i\kappa \right)(t-t')}, \\
 G^A(t,t') \aeq i \theta(t'-t) \me^{-i \left(\omega
 +i\kappa \right)(t-t') }.
\end{eqnarray}
In deriving the above expression, we used the elements of the 
solution (\ref{solution of gamma}) with some algebraic 
manipulations.
For example, 
\bea
{\cal G}(t,t')^{11} \aeq -i \bra 1 \vert 
T [ \gamma(t) \gamma(t')^{\venus} ] \vert 0 \ket 
\nonumber\\
\aeq -i \left\{ \theta(t-t') \bra 1 \vert 
\gamma(t) \gamma^{\venus}(t') \vert 0 \ket 
\right.
\nonumber\\
&& \left. + \theta(t'-t) \bra 1 \vert 
\gamma^{\venus}(t') \gamma(t) \vert 0 \ket \right\}
\nonumber\\
\aeq -i \theta(t-t') \me^{-i(\omega -i\kappa)(t-t')}
\nonumber\\
\aeq G^R(t,t').
\eea
In the third equality, we used (\ref{gamma-gamma}) 
and (\ref{gamma-gamma tilde}).

\subsection{Miscellaneous}

The representation space (the thermal space) of NETFD is
the vector space spanned by the set of bra and ket state
vectors which are generated, respectively, by cyclic
operations of the annihilation operators $\gamma(t)$ and 
$\tilde{\gamma}(t)$ on $\bra 1 \vert$, and of the creation
operators $\gamma^{\venus}(t)$ and
$\tilde{\gamma}^{\venus}(t)$ on $\vert 0 \ket$.

The normal product is defined by means of the annihilation
and the creation operators, i.e.\ $\gamma^{\venus}(t),\ 
\tilde{\gamma}^{\venus}(t)$ stand to the left of
$\gamma(t),\ \tilde{\gamma}(t)$. 
The process, rewriting physical operators in terms of the
annihilation and creation operators, leads to a Wick-type
formula, which in turn leads to Feynman-type diagrams for
multi-point functions in the renormalized interaction
representation.  The internal line in the Feynman-type
diagrams is the unperturbed two-point function 
(\ref{two point func}).

\subsection{Condensation of Particle Pairs}

Introducing the annihilation and creation operators
in the Schr\"odinger representation
\bea
\gamma^{\mu=1} = \gamma_t,\qquad && 
\gamma^{\mu=2} = \tilde{\gamma}^{\venus},
\\
\bar{\gamma}^{\mu=1} = \gamma^{\venus},\qquad  &&
\bar{\gamma}^{\mu=2} = - \tilde{\gamma}_t,
\eea
by the relation
\be
 \gamma(t)^\mu = \hat{V}^{-1}(t) \gamma_t^\mu \hat{V}(t),\qquad 
 \bar{\gamma}(t)^\mu = \hat{V}^{-1}(t) \bar{\gamma}_t^\mu \hat{V}(t),
\label{gamma t}
\ee
with $\hat{V}(t)$ being specified by (\ref{V}),
we can rewrite the hat-Hamiltonian (\ref{b2}) as
\be
 \hat{H} = \omega \left( \gamma^{\venus} \gamma_t - \tilde{\gamma}^{\venus}
 \tilde{\gamma}_t \right)
 - i\hat{\itPi},
\label{H-hat gamma}
\ee
with
\bea
\hat{\itPi} \aeq -\kappa \left( \gamma^{\venus} \gamma_t + \tilde{\gamma}^{\venus}
 \tilde{\gamma}_t + 2 \left[ n(t) - \bar{n} \right] \gamma^{\venus}
 \tilde{\gamma}^{\venus} \right).
\label{Pi-hat gamma}
\eea
It is easily derived by means of the doublet notation
(\ref{hat-H doublet}).

Substituting (\ref{H-hat gamma}) into the quantum master equation
(\ref{b2}), we have 
\bea
 \frac{\partial}{\partial t} \vert0(t)\ket \ \aeq -2 \kappa
\left[ n(t) - \bar{n} \right] \gamma^{\venus}
 \tilde{\gamma}^{\venus} \vert0(t)\ket
\nonumber\\
\aeq \frac{dn(t)}{dt} \gamma^{\venus}  \tilde{\gamma}^{\venus}
\vert 0(t) \ket.
\label{b13}
\eea
It is solved to give
\bea
 \vert0(t)\ket \aeq \exp \left[\int_0^t dt' \frac{dn(t')}{dt'}
\gamma^{\venus}  \tilde{\gamma}^{\venus} \right]
\vert 0 \ket
\nonumber\\
\aeq \exp \left[ \left[n(t)-n(0) \right] \gamma^{\venus} 
 \tilde{\gamma}^{\venus} \right] \vert0\ket.
\label{b3}
\eea
This expression tells us that the time-evolution of 
the unstable vacuum is realized by the condensation of 
$\gamma_k^{\venus}  \tilde{\gamma}_k^{\venus}$-pairs
into the vacuum.
The attractive expression (\ref{b3}), which was obtained
first in \cite{Sudo}, led us to the notion of a mechanism named the {\em 
spontaneous creation of dissipation} 
\cite{guida1,guida2,scd1,scd2,scd3}.
The corresponding {\it order parameter} is given by
\be
\bra 1 \vert \gamma_t \tilde{\gamma}_t \vert 0 \ket 
= -n(t) + n(0)
\label{order parameter}
\ee
where we used the relation
\be
\gamma_t = \gamma_{t=0} - [ n(t) - n(0) ] \tilde{\gamma}^{\venus}.
\ee
We can obtain the results (\ref{b3}) and (\ref{order parameter}) only by algebraic 
manipulations.  This technical convenience of the operator algebra 
in NETFD, which is very much similar to 
that of the usual quantum field theory, enables us to treat open systems in 
far-from-equilibrium state simpler and more transparent 
\cite{Tominaga-Ban,Ban,Tominaga,Iwasaki,Willeboordse,Naoko}.

It also shows that the vacuum is the functional of
the one-particle distribution function $n_k(t)$.
The dependence of the thermal vacuum on $n_k(t)$ is
given by
\be
\frac{\delta}{\delta n_k(t)} \vert 0(t) \ket = 
\gamma_k^{\venus} \tilde{\gamma}_k^{\venus} \vert 0(t) \ket.
\label{vacuum functional}
\ee
We see that the vacuum $\vert 0(t) \ket$ represents the state
where exists the macroscopic object described by 
the one-particle distribution function $n_k(t)$.
The master equation (\ref{master eq})
can be rewritten as
\be
\left\{ \frac{\partial}{\partial t} 
+ \int d^3k\ \frac{d n_k(t)}{d t} \frac{\delta}{\delta n_k(t)} \right\}
\vert 0(t) \ket = 0.
\label{migration of vacuum}
\ee
This shows that the reference vacuum, in this case, is migrating 
in the super-representation space spanned by 
the one-particle distribution function $\{ n_k(t) \}$
with the {\it velocity} $\{ d n_k(t)/dt \}$ as a conserved quantity.

It is easy to see from the normal product form 
(\ref{H-hat gamma}) of \( \hat{H} \) that it satisfies (\ref{left zero}),
since the annihilation and creation operators satisfy
\be
 \gamma_t \vert 0(t) \ket \ = 0,\qquad \bra1\vert \tilde{\gamma}^{\venus} = 0.
\label{b14}
\ee

The hat-Hamiltonian (\ref{b2}) can be also written in the form
\be
 \hat{H} = \omega \left( d^\dagger d - \tilde{d}^\dagger \tilde{d} 
 \right) -i\kappa \left(d^\dagger d + \tilde{d}^\dagger
 \tilde{d} \right),
\label{b4}
\ee
where \( d^{\mu=1} = d,\ d^{\mu=2} = \tilde{d}^\dagger \) and \(
\bar{d}^{\mu=1} = d^\dagger,\ \bar{d}^{\mu=2} = - \tilde{d} \) are 
defined by
\be
 d^\mu = \bar{B}^{\mu\nu} a^\nu,\qquad 
\bar{d}^\mu = \bar{a}^\nu \bar{B}^{-1\nu\mu},
\label{d}
\ee
with
\be
 \bar{B}^{\mu\nu} = \left(
 \begin{array}{cc}
  1+\bar{n} & -\bar{n} \\
  -1 & 1 \\
 \end{array}
 \right).
\ee

The ket-thermal vacuum, \( \vert0\ket \ =\ \vert0(0)\ket  \), 
is specified by (\ref{initial ket}) which can be expressed in terms
of \( d \) and \( \tilde{d}^\dagger \), which are introduced in 
(\ref{d}) below, as
\be
 d \vert0\ket \ = (n-\bar{n})\ \tilde{d}^\dagger \vert0\ket .
\ee

It is easy to see 
from the diagonalized form (\ref{b4}) of \( \hat{H} \) that
\bea
 d(t) \aeq \hat{V}^{-1}(t)\ d\ \hat{V}(t) = d\  e^{-(i \omega +\kappa) t},
\\
 \tilde{d}^{\dd}(t) \aeq \hat{V}^{-1}(t)\ \tilde{d}^\dagger\ \hat{V}(t)
 = \tilde{d}^\dagger\  e^{-(i \omega - \kappa) t}.
\eea

The difference between the operators which diagonalize $\hat{H}$ and
the ones which make $\hat{H}$ in the form of normal product is one
of the features of NETFD, and shows the point 
that the formalism is quite different from usual quantum mechanics and
quantum field theory.  
This is a manifestation of the fact that the
hat-Hamiltonian is a time-evolution generator for
irreversible processes.
In thermal equilibrium state, $n(t) = \bar{n}$, they coincide.

\subsection{Irreversibility}

Let us check here the irreversibility of the system.  
The entropy of the system is given by
\be
 {\cal S}(t) = - \left\{n(t) \ln n(t) - \left[ 1 + n(t) \right]
 \ln \left[ 1 + n(t) \right] \right\},
\label{b7}
\ee
whereas the heat change of the system is given by
\be
 d'Q = \omega dn.
\label{b8}
\ee
Thermodynamics tells us that
\bea
 d{\cal S} = d{\cal S}_e + d{\cal S}_i,&\quad& d{\cal S}_e = d'Q/T_R,\label{b9}\\
 d{\cal S}_i &\geq& 0.
\label{b6}
\eea
The latter inequality (\ref{b6}) is the second law of 
thermodynamics. Putting (\ref{b7}) and (\ref{b8}) into
(\ref{b9}), for \( d{\cal S} \) and \( d{\cal S}_e \), respectively, 
we have a relation for the entropy production rate \cite{suri kagaku}
\be
 \frac{d{\cal S}_i}{dt} = \frac{d{\cal S}}{dt} - \frac{d{\cal S}_e}{dt} 
 = 2 \kappa \left[n(t) - \bar{n} \right] \ln 
 \frac{n(t)[1 + \bar{n}]}{\bar{n}[1 + n(t)]}\label{b10}
 \geq 0.
\label{b11}
\ee
It is easy to check that the expression on the right-hand side of the second
equality satisfies the last inequality which is consistent
with (\ref{b6}). 
The equality realizes either 
for the thermal equilibrium state, \( n(t) = \bar{n} \), or for 
the quasi-stationary process, \( \kappa \to 0 \).

\section{Ito and Stratonovich Multiplications}
\label{Multiplication}

The definitions of the Ito \cite{Ito} and the Stratonovich 
\cite{Strat} multiplications are given, respectively, by
\bea
\lefteqn{
X^{(H)}(t) \cdot dY^{(H)}(t) 
}\nonumber\\
\aeq X^{(H)}(t) \left[ Y^{(H)}(t + dt) 
- Y^{(H)}(t) 
	\right],
	\label{def-Ito-1}\\
\lefteqn{
dX^{(H)}(t) \cdot Y^{(H)}(t) 
}\nonumber\\
\aeq \left[ X^{(H)}(t + dt) - X^{(H)}(t) 
	\right] Y^{(H)}(t),
	\label{def-Ito-2}
\eea
and
\bea
X^{(H)}(t) \circ dY^{(H)}(t) 
\aeq \frac{X^{(H)}(t + dt) + X^{(H)}(t)}{2}
\nonumber\\
&&	\left[ Y^{(H)}(t + dt) - Y^{(H)}(t) \right],
	\label{def-Strat-1}\\
dX^{(H)}(t) \circ Y^{(H)}(t) 
\aeq \left[ X^{(H)}(t + dt) - X^{(H)}(t) 
	\right]
\nonumber\\
&&	\frac{Y^{(H)}(t + dt) + Y^{(H)}(t)}{2},
	\label{def-Strat-2}
\eea
for arbitrary stochastic operators $X^{(H)}(t)$ and
$Y^{(H)}(t)$ in the Heisenberg representation.
From (\ref{def-Ito-1}), (\ref{def-Ito-2}) and (\ref{def-Strat-1}),
(\ref{def-Strat-2}), we have the formulae
which connect the Ito and the Stratonovich products in the
differential form
\bea
X^{(H)}(t) \circ dY^{(H)}(t) \aeq X^{(H)}(t) dY^{(H)}(t) 
\nonumber\\
	&& + \frac12 dX^{(H)}(t) \cdot dY^{(H)}(t),
	\label{connect-1}\\
dX^{(H)}(t) \circ Y^{(H)}(t) \aeq dX^{(H)}(t) \cdot Y^{(H)}(t) 
\nonumber\\
	&& + \frac12 dX^{(H)}(t) \cdot dY^{(H)}(t).
	\label{connect-2}
\eea

The connection formulae for the stochastic operators in the Schr\"odinger
representation are given, in the same form as (\ref{connect-1})
and (\ref{connect-2}), by
\bea
X^{(S)}(t) \circ dY^{(S)}(t) \aeq X^{(S)}(t) dY^{(S)}(t) 
\nonumber\\
	&& + \frac12 dX^{(S)}(t) \cdot dY^{(S)}(t),
	\label{connect-S-1}\\
dX^{(S)}(t) \circ Y^{(S)}(t) \aeq dX^{(S)}(t) \cdot Y^{(S)}(t)
\nonumber\\
	&& + \frac12 dX^{(S)}(t) \cdot dY^{(S)}(t),
	\label{connect-S-2}
\eea
where the operators $X^{(S)}(t)$ and $dX^{(S)}(t)$ in the 
Schr\"odinger representation are introduced respectively through 
$X^{(H)}(t) = \hat{V}_f^{-1}(t) X^{(S)}(t) \hat{V}_f(t)$ and 
$dX^{(H)}(t) = \hat{V}_f^{-1}(t) dX^{(S)}(t) \hat{V}_f(t)$.

\section{Quantum Brownian Motion} \label{q Brownian motion}

Let us introduce the annihilation and creation operators 
$b_t$, $b^\dagger_t$ and their tilde conjugates 
satisfying the canonical commutation relation:
\be
[ b_t,\ b^\dagger_{t'} ] = \delta(t - t'),\quad 
[ \tilde{b}_t,\ \tilde{b}^\dagger_{t'} ] = \delta(t - t').
\ee
The vacuums $( \vert$ and $\vert )$ are defined by 
\be
b_t \vert ) = 0,\quad \tilde{b}_t \vert ) = 0,\quad 
( \vert b^\dagger_t = ( \vert \tilde{b}_t.
\ee
The argument $t$ represents time.

Introducing the operators
\bea
B_t \aeq \int_0^{t-dt} dB_{t'} = \int_0^t dt'\ b_{t'},\\
B^\dagger_t \aeq \int_0^{t-dt} dB^\dagger_{t'} 
= \int_0^t dt'\ b^\dagger_{t'},
\label{B}
\eea
and their tilde conjugates for $t \geq 0$, we see that they satisfy
$
B(0) = 0
$, 
$
B^\dagger(0)=0
$, 
\be
[B_s,\ B^\dagger_t ] = \mbox{min}(s,t),
\label{commutation B}
\ee
and their tilde conjugates, and that 
they annihilate the vacuum $\vert )$ with the thermal state condition for 
$( \vert$:
\be
dB_t \vert ) = 0,\quad d\tilde{B}_t \vert ) = 0,\quad
( \vert d B^\dagger_t = ( \vert d\tilde{B}_t.
\ee
These operators represent the quantum Brownian motion.

Let us introduce a set of new operators by the relation
\be
dC_t^\mu = \bar{B}^{\mu \nu} dB_t^\nu,
\ee
with the Bogoliubov transformation defined by
\bea
 \bar{B}^{\mu\nu} = \left(
  \begin{array}{cc}
   1+ \bar{n} & -\bar{n} \\
   -1 & 1 \\
  \end{array}
 \right),
\label{B bar}
\eea
where $\bar{n}$ is the Planck distribution function.
We introduced the thermal doublet:
\bea
dB_t^{\mu=1} = dB_t, && \quad dB_t^{\mu=2} = d\tilde{B}^\dagger_t,
\\
d\bar{B}_t^{\mu =1} = dB^\dagger_t, && \quad d\bar{B}_t^{\mu = 2} = -
d\tilde{B}_t,
\eea
and the similar doublet notations for $dC_t^\mu$ and 
$d{\bar C}_t^\mu$.
The new operators annihilate the new vacuum $\bra \vert$, and 
have the thermal state condition for $\vert \ket$:
\be
dC_t \vert  \ket = 0,\quad d\tilde{C}_t \vert  \ket = 0, 
\quad 
\bra  \vert d C^\dagger_t = \bra \vert 
d\tilde{C}_t.
\label{cal B bra}
\ee

We will use the representation space constructed on
the vacuums $\bra \vert$ and $\vert \ket$.  
Then, we have, for example,
\bea
\bra \vert dB_t \vert \ket \aeq \bra \vert dB^\dagger_t \vert \ket = 0,\\
\bra \vert dB^\dagger_t dB_t \vert \ket = \bar{n} dt,
&&
\bra \vert dB_t dB^\dagger_t \vert \ket 
= \left( 1+\bar{n} \right) dt.
\label{B-Bd}
\eea
They can be written 
\bea
dB_t^\dagger dB_t \aeq dB_t d\tilde{B}_t = \bar{n} dt,
\\
dB_t dB_t^\dagger \aeq dB_t^\dagger d\tilde{B}_t^\dagger 
= (1+\bar{n}) dt,
\eea
as weak relations.

%%%%%%%%%%%%%%%%%%%%%%%%%%%%%%%%%%%%%%%%%%%%%%%%%%%%%%%%%%%%%%%%%
% References:
%%%%%%%%%%%%%%%%%%%%%%%%%%%%%%%%%%%%%%%%%%%%%%%%%%%%%%%%%%%%%%%%%

\end{document}